\documentclass[preprint,12pt,a4paper]{elsarticle}

\usepackage{amssymb}
\usepackage{float}
\usepackage{listings}
\PassOptionsToPackage{hyphens}{url}
\usepackage{hyperref}
\usepackage{xcolor}
\usepackage{lineno}
\lstdefinestyle{mystyle}{
    basicstyle        = \ttfamily\footnotesize,
    breakatwhitespace = false,
    breaklines        = true,
    captionpos        = b,
    keepspaces        = true,
    numbersep         = 5pt,
    showspaces        = false,
    showstringspaces  = false,
    showtabs          = false,
    keywordstyle      = \color{blue},
    stringstyle       = \color{green},
    commentstyle      = \color{red}\ttfamily
}
\restylefloat{table}
\journal{SoftwareX}

\begin{document}

\begin{frontmatter}

\title{Omnisolver: An extensible interface to Ising spin--glass and QUBO solvers: adding a distributed GPU brute-force plugin}

\author[1]{Konrad Ja\l{}owiecki}
\cortext[cor1]{Corresponding author}
\author[2,3]{Jakub Paw{\l}owski}
\author[1,2]{Bart{\l}omiej Gardas}
\author[1,2]{{\L}ukasz Pawela\corref{cor1}}
\ead{lpawela@iitis.pl}

\address[1]{Institute of Theoretical and Applied Informatics, Polish Academy
of Sciences, Ba{\l}tycka 5, 44-100 Gliwice, Poland}

\address[2]{Quantumz.io Sp. z o. o., Pu{\l}awska 12/3, 02-566 Warsaw, Poland}

\address[3]{Institute of Theoretical Physics, Faculty of Fundamental Problems
of Technology, Wroc{\l}aw University of Science and Technology,
50-370 Wroc{\l}aw, Poland}

\begin{abstract}
This software update extends
Omnisolver~\cite{jalowiecki2021omnisolver} with
\texttt{omnisolver-bruteforce}, a first-class plugin for exact exhaustive
search of QUBO and Ising instances on CUDA-enabled GPUs. The update
contributes three components: packaging of the single-GPU brute-force
kernel of~\cite{jalowiecki2021brute} as a first-class Omnisolver plugin
with a uniform Python and
CLI interface, a Ray-based distributed framework that splits the search
into $2^{k}$ fixed-variable subproblems and dispatches them across multiple
GPUs and hosts, with a controller that gathers and merges partial results,
and numerical stabilization of the fast \texttt{float32} ground-state
path through compensated incremental updates, periodic exact energy
re-anchoring, and best-buffer refresh. Stabilization activates automatically
for $N \geq 40$ leaving the public sampler API is unchanged. On dense random Ising
instances we measure exhaustive solve times up to $N=60$ on $8\times$ NVIDIA
H100 (96\,GB) GPUs, reaching $\approx\!3.15$ days at $N=60$ in close
agreement with the empirical $t_{N+1}=2\,t_N$ doubling rule, with the
distributed sampler reaching the full $\approx\!8\times$ speedup over a
single H100 from $N\!\gtrsim\!44$ onwards. The plugin thereby serves as a
practical ground-truth oracle for state-of-the-art classical heuristics such
as the Simulated Bifurcation Machine.
\end{abstract}


\end{frontmatter}

\section*{Refers to}
K.~Ja\l{}owiecki, {\L}.~Pawela, ``Omnisolver: An extensible interface to Ising
spin--glass and QUBO solvers,'' \emph{SoftwareX} \textbf{24}, 101559 (2023).
DOI: \url{https://doi.org/10.1016/j.softx.2023.101559}.

\section*{Metadata}
\label{metadata}

\begin{table}[H]
\setlength{\tabcolsep}{4pt}
\begin{tabular}{|l|p{5.6cm}|p{6.4cm}|}
\hline
\textbf{Nr.} & \textbf{Code metadata description} & \textbf{Value} \\
\hline
C1 & Current code version & 0.0.5 \\
\hline
C2 & Permanent link to code/repository used for this code version &
\url{https://github.com/euro-hpc-pl/omnisolver-bruteforce/releases/tag/0.0.5} \\
\hline
C3 & Legal Code License & Apache License 2.0 \\
\hline
C4 & Code versioning system used & git \\
\hline
C5 & Software code languages, tools, and services used & Python, C++, Cython,
CUDA, Ray \\
\hline
C6 & Compilation requirements, operating environments \& dependencies & Linux,
CUDA Toolkit, \texttt{Python >= 3.10}, \texttt{omnisolver >= 0.0.4},
\texttt{dimod}, \texttt{numba}, \texttt{numpy}, \texttt{ray} \\
\hline
C7 & Link to developer documentation/manual &
\url{https://euro-hpc-pl.github.io/omnisolver/} \\
\hline
C8 & Support email for questions & \verb|lpawela@iitis.pl| \\
\hline
\end{tabular}
\caption{Code metadata}
\label{codeMetadata}
\end{table}


\section{Description of the software update}
\label{sec:description}

Omnisolver~\cite{jalowiecki2021omnisolver} is a plugin-based framework that
exposes a uniform Python and command-line interface to QUBO/Ising solvers,
integrated with \texttt{dimod}. The original publication established the
framework and its first plugins, for example, the Parallel Tempering solver. This update introduces
\texttt{omnisolver-bruteforce}, a dedicated plugin that promotes a previously
stand-alone single-GPU exhaustive solver~\cite{jalowiecki2021brute} to a
first-class Omnisolver component, adds distributed multi-GPU execution
through Ray~\cite{DBLP:journals/corr/abs-1712-05889}, and stabilizes the
numerical behavior of its fast ground-state path.

\paragraph{Purpose of an exhaustive solver in Omnisolver}
An exhaustive backend provides \emph{certified} reference, i.e. lowest-energy,
solutions for QUBO/Ising instances. Such ground-truth answers are essential for
benchmarking approximate or heuristic methods, for verifying NISQ-oriented
workflows~\cite{preskill2018quantum}, and for assessing claims of quantum
runtime advantage. In recent years, the latter point has become especially
critical in the community, with many studies relying on metrics that
explicitly depend on the ground-state energy, such as the \emph{optimality gap}.
For example, in recent comparative studies of runtime
advantage~\cite{tuziemski2026limits} and of the quantum-classical scaling
gap~\cite{pawlowski2025closing}, quantum resources were benchmarked against the
Simulated Bifurcation Machine
(SBM)~\cite{goto2016bifurcation,goto2019combinatorial,goto2021high,pawlowski2026sbqa},
the current state-of-the-art classical heuristic for dense QUBO/Ising problems.
Such results rely on the ability to certify the true optimum on representative
instances, which is precisely what an exhaustive GPU solver provides.

\begin{figure}[t]
\centering
\includegraphics[width=0.95\textwidth]{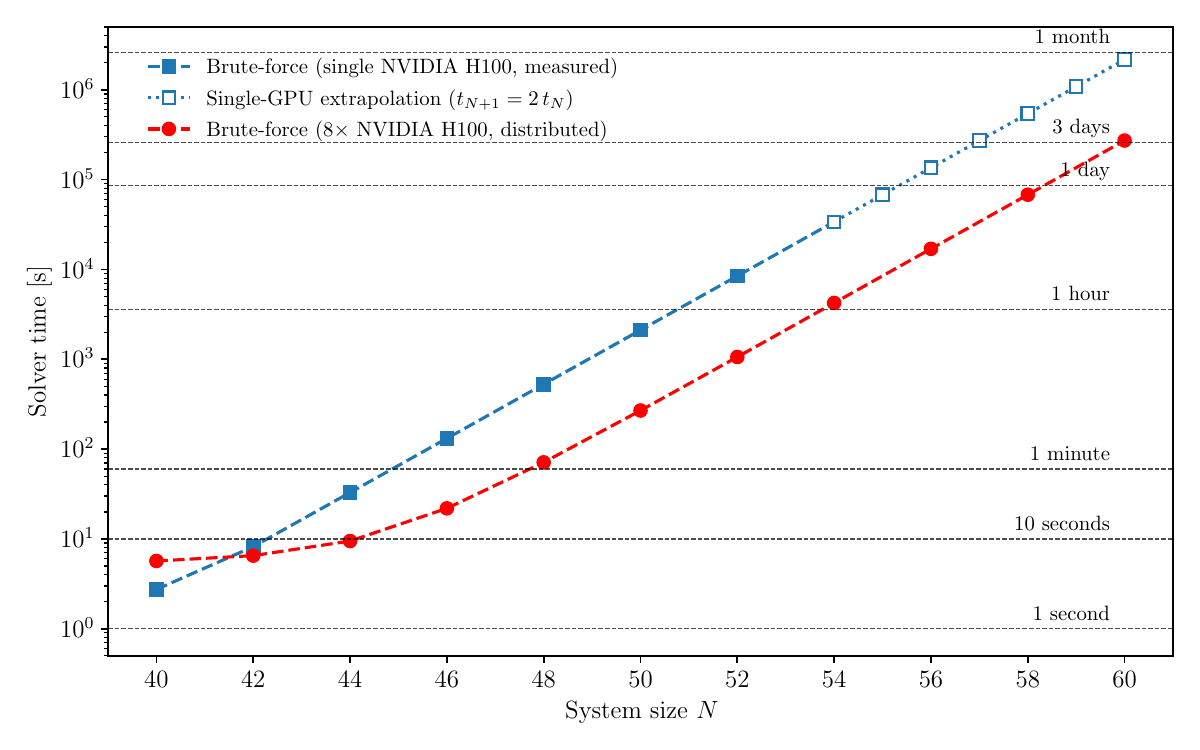}
\caption{Wall-clock runtime of \texttt{omnisolver-bruteforce} on all-to-all
    random Ising instances with $J_{ij},h_i\sim \mathcal{U}[-1,1]$, measured
    on a local in-house cluster of $8\times$ NVIDIA H100 (96\,GB) GPUs (two
    nodes $\times$ 4 GPUs). Each filled marker is a single solver run on
    one fixed instance per system size $N$. Blue squares: single-GPU
    sampler (one of the H100s, measured for $N=40,\dots,54$). Open blue
    squares: doubling-rule extrapolation $t_{N+1}=2\,t_N$ from the largest
    measured single-GPU point to $N=60$. Red circles: distributed sampler
    on all eight GPUs (measured for $N=40,\dots,60$). The two curves
    cross between $N=40$ and $N=42$; from $N\!\gtrsim\!44$ the
    distributed-to-single-GPU ratio settles at the linear-scaling value
    $\approx\!8$.}
\label{fig:runtimes}
\end{figure}

\paragraph{Architecture}
The plugin introduced by this software update exposes two sampler classes:
\texttt{BruteforceGPUSampler} (single-node, single-GPU) and
\texttt{DistributedBruteforceGPUSampler} (multi-GPU/multi-node via Ray). The
distributed sampler fixes \texttt{num\_fixed\_vars}\,$=k$ variables and
enumerates all $2^{k}$ fixed assignments, each of which becomes a subproblem
with $N-k$ unfixed variables, solved by the GPU kernel on a worker. Partial
solutions are merged on the controller node. As a practical guideline, the
single-GPU sampler is the right choice while one device has both the memory
budget and a wall-clock time within reason for the chosen problem size;
beyond that point, the distributed sampler amortizes the work by powers of
two in the number of workers. On our hardware, it crosses the break-even
region between $N=40$ and $N=42$ (see Fig.~\ref{fig:runtimes}), and from
$N\!\gtrsim\!44$ it reaches the full $\approx\!8\times$ speedup expected
from the 8-GPU pool, with Ray-scheduling and kernel-launch overheads
amortized over the per-subproblem compute. Performance-critical
kernels are implemented in CUDA/C++ and exposed to Python via Cython; the
\texttt{num\_states=1} fast path operates in \texttt{float32} to maximize GPU
throughput, with the stabilization mechanisms described below compensating
for the resulting roundoff.

\paragraph{Relation to the predecessor single-GPU solver}
On instances that fit a single device, the kernel is API-compatible with
the original solver of~\cite{jalowiecki2021brute} and reproduces its
per-state throughput within measurement noise; the present update does not
re-benchmark the single-device axis but inherits its performance envelope
and adds, on top of it, the stabilization patches and the distributed
controller.

\paragraph{Update contributions}
\begin{itemize}
    \item \textbf{First-class Omnisolver plugin}: the brute-force kernel is
    available through Omnisolver's plugin mechanism with a uniform Python and
    CLI interface, returning \texttt{dimod.SampleSet} with runtime metadata.
    \item \textbf{Distributed multi-GPU execution} via a Ray-based fixed-variable
    decomposition, enabling exhaustive search at sizes infeasible on a single
    device within a practical time-frame.
    \item \textbf{Stabilized fast ground-state path}: compensated incremental
    updates, periodic exact energy re-anchoring, and best-buffer refresh
    eliminate the \texttt{float32} drift previously observed for $N \geq 40$
    in the \texttt{num\_states=1} mode. Stabilization activates automatically;
    the API is unchanged.
    \item \textbf{Hardening for large-$N$ enumeration}: 64-bit-safe bit
    operations and kernel-step bookkeeping support spin configurations of
    sizes well beyond the original single-GPU regime.
    \item \textbf{Extended validation}: GPU tests cross-check energies against
    from-scratch exact recomputation and against \texttt{dimod.ExactSolver} on
    tractable instances.
\end{itemize}

\paragraph{Performance}
Figure~\ref{fig:runtimes} reports wall-clock solve times of both sampler
modes on all-to-all random Ising instances with $J_{ij},h_i\sim
\mathcal{U}[-1,1]$ i.i.d., generated by the benchmark script
\texttt{code/bf.py} from the article repository (one instance per size,
fixed seed); measured on a local in-house cluster of $8\times$ NVIDIA
H100 (96\,GB) GPUs (two nodes, 4 GPUs per node, interconnected by standard
Ethernet). The single-GPU sampler was run on one of the same H100s, the
distributed sampler on all eight. The $O(2^N)$ doubling per added spin
is the dominant feature on both curves and is clean already at $N=40$ on
the single-GPU axis; the distributed curve flattens out for $N \lesssim 44$
because timings are dominated there by kernel-launch and Ray-scheduling
overhead. The distributed sampler crosses below the single-GPU sampler
between $N=40$ and $N=42$, and reaches the full $\approx\!8\times$ speedup
(matching the 8-GPU pool) from $N\!\gtrsim\!44$ onwards. The largest
measured size, $N=60$ in the distributed mode, takes
$\approx\!272{,}300$ s ($\approx 3.15$ days), within $0.3\%$ of the
doubling-rule extrapolation from $N=58$ ($2^{2}\!\times\!67{,}900 \approx
271{,}600$ s); this empirically confirms the $t_{N+1}=2\,t_N$ rule across
the measured range. We use the same rule to project the single-GPU
sampler (measured up to $N=54$) to $N=60$ along the dashed extension in
Fig.~\ref{fig:runtimes}: the projected single-GPU wall-clock at $N=60$ is
$\approx\!2.17\times 10^{6}$ s ($\approx\!25$ days), about $8.0\times$
the measured 8-GPU time at the same size and consistent with linear
strong scaling. The same rule places $N=64$ at $\approx\!50$ days on the
8-GPU pool. Constant-factor deviations from exact doubling (kernel-launch
overhead, host/device transfer, Ray scheduling) can shift the absolute
numbers by at most a few percent but not the asymptotic scaling.

\paragraph{Scaling outlook on larger allocations}
The fixed-variable decomposition is embarrassingly parallel: each of the
$2^k$ fixed assignments defines an independent GPU subproblem, with no
inter-worker communication during the search and only a trivial merge step
on the controller. As a back-of-the-envelope projection, scaling our 8-GPU
baseline to $2^{16}$ GPUs, an allocation within reach of
leadership-class systems, with the natural choice $k=16$ placing one
subproblem per worker, would reduce $N=60$ from $\approx\!3.15$ days
to $\approx\!33$ s, and the projected $N=64$ time from $\approx\!50$
days to $\approx\!9$ min. Under a one-month wall-clock budget on the same
$2^{16}$-GPU allocation, the doubling rule places the limit for certifiable exhaustive
search at $N \lesssim 76$. This linear-in-GPU-count outlook is consistent
with the algorithmic structure; however, in practice, controller-side merge
bandwidth, Ray scheduling overhead, and per-kernel launch latency would set
the true scaling ceiling, especially as $2^k$ approaches the number of
available workers.

\paragraph{Precision of the reported energies}
Because the GPU backend operates in \texttt{float32} on the fast ground-state
path, returned energies are subject to incremental-update roundoff. We
re-evaluate the energy of each returned spin configuration from scratch in
\texttt{float64}. For the largest measured instances ($N=56,58,60$), the
absolute deviation between reported and recomputed energies stays below
$4\times 10^{-6}$ for energies of magnitude $\mathcal{O}(10^{2})$,
roughly eight significant digits, consistent with the expected accuracy of
the stabilized \texttt{float32} accumulator. Using the same recomputation,
we cross-check an SBM run on the same instances: a 
simulated bifurcation solver running on a single H100 GPU with
$2^{12}=4096$ parallel samples and $3{,}000$ integration steps, reporting
the lowest-energy sample. The recomputed energies of the brute-force and
SBM solutions agree to within $6\times 10^{-6}$, and the returned spin
configurations are bit-identical (Hamming distance $0$) at $N=56$, $58$,
and $60$. On these instances, SBM reaches the exact ground state certified
by exhaustive search, and the residual energy-value differences are
explained entirely by single-precision roundoff. The full verification
harness (Julia driver, SBM parameters, and the resulting
\texttt{bf\_velox\_verification\_*.csv} tables) is available in the
companion repository
\href{https://github.com/euro-hpc-pl/omni-bench}{\texttt{euro-hpc-pl/omni-bench}},
together with all \texttt{instances/*.txt} and \texttt{results/*.json}
files that produced Fig.~\ref{fig:runtimes}.

\paragraph{Limitations and intended scope}
Exhaustive enumeration scales as $O(2^N)$ irrespective of GPU count or
numerical precision: the update lowers the constant factor and removes a
\texttt{float32} drift pitfall, but it cannot change the underlying
complexity. The plugin is therefore intended as a \emph{verification and
certification backend} for QUBO/Ising research pipelines---providing
ground-truth optima against which heuristics, annealers, and NISQ-era
quantum approaches can be calibrated---rather than as a routine optimizer
for application-scale problems, where heuristic solvers such as the SBM
remain the appropriate primary tool. Concrete hard limits on the current
implementation are: (i) the spin index is held in a 64-bit word, capping
the supported problem size at $N \leq 64$; (ii) the per-worker
\texttt{suffix\_size} (the number of variables enumerated inside one CUDA
kernel launch) is bounded by the GPU's L2/working-set budget and is set to
$23$ on H100\,/\,96\,GB in our runs, so larger $N$ must be reached by
raising \texttt{num\_fixed\_vars}, not \texttt{suffix\_size}; (iii) the
projected runtimes for $N > 60$ are back-of-the-envelope extrapolations
under the empirical $t_{N+1}=2\,t_N$ rule and may drift by a small constant
factor on different stacks. Reported runtimes
additionally depend on the hardware and software stack (GPU model, CUDA
toolkit, Ray version, interconnect), so the absolute timings in
Fig.~\ref{fig:runtimes} should be read as a hardware-specific operating
point rather than a universal guarantee.

\section{Illustrative example}
\label{sec:example}
The plugin is distributed as \texttt{omnisolver-bruteforce} on PyPI
(\texttt{pip install omnisolver-bruteforce}) and is usable either
stand-alone or through the Omnisolver CLI as the \texttt{bruteforce-gpu}
solver. Complete documentation and the benchmark script used to produce
Fig.~\ref{fig:runtimes} are available in the
\href{https://github.com/euro-hpc-pl/omnisolver-bruteforce}{GitHub
repository}~(C2).

A minimal distributed-sampler invocation looks as follows:

\begin{lstlisting}[language=Python,caption={Minimal distributed brute-force example},label={lst:bf-min}]
from omnisolver.bruteforce.gpu.distributed import DistributedBruteforceGPUSampler
from dimod.serialization import coo

with open("instance.txt") as fd:
    bqm = coo.load(fd, vartype="SPIN")

sampler = DistributedBruteforceGPUSampler()
result = sampler.sample(
    bqm,
    num_states=1,
    num_fixed_vars=3,
    suffix_size=23,
    grid_size=8192,
    block_size=1024,
    partial_diff_buffer_depth=10,
    num_steps_per_kernel=8192,
)
print(result.first.energy, result.info["solve_time_in_seconds"])
\end{lstlisting}

For multi-GPU execution, Ray must be started on the head and worker
nodes (here, an 8-GPU two-node setup with 4 GPUs per node), after which the
benchmark script reproduces the curve in Fig.~\ref{fig:runtimes}:

\begin{lstlisting}[language=bash,caption={Ray startup and benchmark command used to produce Fig.~\ref{fig:runtimes}},label={lst:ray-commands}]
# Head node
ray stop --force
ray start --head --port=6379 --num-gpus=4

# Worker node
ray stop --force
ray start --address='<HEAD_NODE_IP>:6379' --num-gpus=4

# Run benchmark sweep (from the article repository root)
python code/bf.py --start 40 --stop 60 --step 2 \
                  --sampler-mode distributed --seed 42 \
                  --skip-existing
\end{lstlisting}

\section{Conclusions and outlook}
\label{sec:conclusions}
This update promotes the CUDA brute-force kernel
of~\cite{jalowiecki2021brute} to a first-class Omnisolver plugin, equips it
with distributed multi-GPU execution through Ray, and resolves the
\texttt{float32} stability issue of the fast ground-state path---all
without changing the public sampler API. On dense random Ising instances we
now certify exhaustive ground states up to $N=60$ on $8\times$ H100 GPUs,
empirically confirming the $t_{N+1}=2\,t_N$ doubling rule across the
measured range. Natural next steps are: (i) exposing the re-anchoring
cadence and compensated-update mode as user-tunable knobs for instance
families where the default policy is suboptimal; (ii) replacing the
controller-side merge by a hierarchical reduction so the algorithm scales
cleanly to $2^k$ workers with $k$ approaching the system size; and (iii)
publishing a versioned reproducibility capsule (fixed seeds, hardware
descriptors, command presets) so that runtime and numerical claims can be
re-checked on heterogeneous clusters.

\section*{Declaration of competing interest}
The authors declare that they have no known competing financial interests or
personal relationships that could have appeared to influence the work reported
in this paper.

\section*{Acknowledgements}
This work was supported by the National Science Center (NCN), Poland, via
project Sonata Bis 10, No.~2020/38/E/ST3/00269 (B.G.) and Sonata Bis 15,
No.~2025/58/E/ST6/00422 ({\L}.P.). Quantumz.io Sp.~z~o.o.\ acknowledges
support received from the Polish Agency for Enterprise Development (PARP),
Poland, under Project No.~FENG.01.01-IP.02-0625/23, titled
\textit{Dynamic allocation of resources in industrial ecosystems susceptible
to disturbances using physics-inspired algorithms and machine learning}.

\bibliographystyle{elsarticle-num}
\bibliography{bruteforce.bib}

\end{document}